\def\tr{\hbox{tr}}
\def\ln{\ell{n}}

\documentstyle[12pt,aps]{revtex}

\begin{document}
\draft
\title{Chiral gauged fermions on a lattice}
\author{She-Sheng Xue
}
\address{
ICRA, INFN  and
Physics Department, University of Rome ``La Sapienza", 00185 Rome, Italy
}


\maketitle

\centerline{xue@icra.it}

\begin{abstract}

The chiral fermion model with local multifermion interactions proposed in Nucl.~Phys.~B486 (1997) 282 and Phys.~Rev.~D61 (2000) 054502 processes an exact
$SU_L(2)$ chiral gauge symmetry and $SU_L(2)\otimes U_R(1)$ chiral flavour symmetry on a lattice and a plausible scaling region for the target chiral gauge theory in the continuum limit. Following the previous analysis of massive and massless fermion spectra in the scaling region, we compute the one-particle-irreducible coupling vertices between gauge field and fermions by the strong multifermion coupling expansion and analytical continuation of these vertex functions in the momentum space. We show a peculiar scenario that a massless fermion is $SU_L(2)$-chirally gauged in the low energy and 15 non-degenerate massive Dirac fermions are $SU_L(2)$-vectorially gauged at the lattice scale $O(1/a)$. The Ward identities associated to the chiral gauge symmetry are realized by both the massless chiral fermion and massive Dirac fermions. These Ward identities protect the perturbative calculations in the small gauge coupling from hard gauge-symmetry breakings and lead to the normal gauge-invariant renormalization prescription. The vacuum functional is perturbatively computed by a continuum regularization scheme in 16 edges of Brillouin zones. We achieve the correct form of the gauge anomaly and $U_L(1)$ fermion-flavour singlet anomaly with the soft chiral symmetry breaking scale that is much smaller than the lattice scale. The residual breakings of chiral gauge symmetry after the gauge anomaly cancellation are eliminated in the normal gauge-invarinant renormalization prescription. We discuss the consistency of the scenario and the reasons for it to work for perturbative and non-perturbative gauge field.

\end{abstract}

\pacs{
11.15Ha,
11.30.Rd, 
11.30.Qc
}

\narrowtext

\section{Introduction}

The problems of chiral gauge theories on a lattice concerning  the vectorlike (doubling) 
phenomenon, chiral gauge symmetries and  anomalies have been still bothering 
theoretical physicists, since the ``no-go'' theorem \cite{nn81} of Nielsen and Ninomiya was
demonstrated in 1981.  However, many progresses in understanding  
have been made for approaching to the solution of the problem\cite{ep}-\cite{testa99}. These 
approaches can be very briefly categorized into two major classes, one is the modeling 
based on appropriately introducing local interactions, another is the modeling, without 
interactions, instead based on delicately manipulating chiral fermions and 
gauge fields on the lattice. The latter has been quickly developed as the most 
exciting approaches with the strong theoretical ground\cite{nerev,lurev,gs99}. While as for the modeling with interactions, it has become a general belief that such modeling cannot lead to any chiral 
gauge theories on the lattice. Nevertheless, following the studies of the 
multifermion coupling model we proposed in refs.\cite{xue97,xue97l,xue99}, we are attempt to discuss the aspects of the coupling between gauge field and fermion fields, the gauge anomaly, fermion-flavour singlet anomaly (fermion-number violation) and the elimination of residual gauge symmetry breakings in this paper. 

The main idea for modeling with interactions , either multifermion couplings (see examples \cite{ep,px91,mc,xue97}) or scale-fermion couplings (see example \cite{ss,ahx,monrev}) is  
analogous, although the details of interactions can be 
very different. It is to use strong local interactions to gauge-invariantly decouple 
extra chiral fermion species and in the meantime obtain the correct gauge anomalies 
and fermion-number violation in the continuum limit. The Eichten-Preskill\cite{ep} and 
Smit-Swift model\cite{ss,ahx} were two of such models. The studies of these two models 
show a broken phase with a hard spontaneous symmetry breaking of Nambu-Jona-Lasinio type\cite{njl} at the lattice scale ($O(1/a)$). 
This broken phase separates two chiral gauge symmetric phases\cite{perev,pgr} in the phase 
diagram in terms of couplings. This hard spontaneous symmetry breaking enforces the masses of intermediate gauge bosons to be at the lattice scale ($O(1/a)$)\cite{ahx,pxm}. On the other hand, even within the 
chiral gauge symmetric phase, due to the exact locality of interactions introduced and the argument of anomalies cancelation, 
the gauge symmetric spectrum of fermion zero modes must be vectorlike 
analogously to the context of the ``no-go" theorem. These
studies lead to the general belief that the chiral gauge symmetry realized by a chiral fermion spectrum in the low-energy cannot be maintained on the lattice. Even though we do not share the view of this general belief, these studies tell us that in such models with interactions, the hard spontaneous symmetry breaking  at the lattice scale absolutely cannot be tolerated in order to have a loophole to achieve a correct chiral gauge theory on the lattice. In the model of multifermion couplings with an anomaly-free fermion content, for many years\cite{px91,pxm,px}, we have been trying to find the appropriate multifermion couplings\cite{xue97} that can avoid the hard spontaneous symmetry breaking within a scaling region in the phase space, and in this scaling region composite particles with ``wrong" chirality can dissolve to their cuts\cite{xue97l}, to which the ``no-go" theorem does not apply, so as to preserve exact chiral gauge symmetries with the chiral fermion spectrum in the continuum limit\cite{xue99}. Due to the complex 
dynamics of strong interactions of such models that belong to various different universality classes, it is difficult to demonstrate a model with peculiar interactions to 
work, but it even much more difficult to prove a general ``no-go" theorem for the failure of any models so 
constructed.  The recent development of lattice chiral gauge theories\cite{nerev,lurev} 
shows that the chiral gauge symmetries can 
be preserved on the lattice, and this shines a light on this great puzzle whether chiral 
gauge symmetries realized by the chiral fermion spectrum can be achieved. 

The essential spirit of the ``no-go'' theorem of Nielsen and Ninomiya is that, 
under very general prerequisites, e.g., locality and gauge symmetry,  
which are usually required in quantum field theories,
the paradox concerning chiral gauge symmetries, vectorlike doubling and
anomalies are unavoidably entwined. In this paper, within the context of the multifermion coupling 
model and its scaling region we proposed in refs.\cite{xue97,xue99}, where no any hard spontaneous symmetry breaking occurs, we will discuss the chiral fermion spectrum in the low energy, vectorlike fermion spectrum in the high energy and their coupling to the gauge field to show how chiral gauge 
symmetry is realized. In addition, we will show the computations of the correct gauge anomaly 
and fermion-flavour singlet anomaly and discuss the reasons for such anomalies arise in the scaling region of our model.
The elimination of residual breakings of the gauge symmetry after concelation of the gauge anomaly is discussed. 

The paper 
is organized as follow. In section II, we give a detail review of the properties of our model in the scaling region, which were  obtained in Ref.\cite{xue97,xue97l,xue99}. In section III, we demonstrate the exact decoupling of the right-handed chiral fermion from the gauge field by the exact shift-symmetry of the right-handed chiral fermion in our model. This plays an  extremely important r\^ole in obtaining the correct fermion-flavour singlet anomaly. In section IV, based on (i) the recursion relations obeyed by the three-point Green functions of the gauge and fermion fields in the strong coupling limit; (ii) the analytical continuation of these Green functions in the energy-momentum space; (iii)  the weak-coupling computations of these Green functions, we show the exact chiral gauge symmetry and associated Ward identities are realized by both chiral fermion spectra in low-energy region and vectorlike fermion spectra in high energy regions. In section V, we present an explicit perturbative computation for achieving the correct gauge anomaly and discuss how the correct gauge anomaly arises from a soft chiral symmetry breaking at the scale much smaller than the lattice scale. We discuss the cancelation of the gauge anomalies and elimination of residual chiral gauge symmetry breaking. These are important for our model to have a consistent scenario of preserving chiral gauge symmetries with the correct low-energy spectrum and to have the normal renormalization prescription in perturbative calculations of the small gauge coupling. Some preliminary results of sections IV and V were reported in Phys. Lett. B408 (1997) 299. In section VI, we explicitly derive the 
fermion-flavour singlet anomaly via the mixing anomaly due to a soft chiral symmetry breaking and we discuss how this fermion-flavour singlet anomaly is obtained from the action (\ref{action}), which at the lattice scale is global fermion flavour symmetric. In the conluding section, 
we also present a very general and brief discussion and remark on the possible connection to the recent development of lattice chiral gauge theories based on the Ginsparg-Wilson relation \cite{gw}. 

\section{A plausible scaling region of lattice chiral fermions}

We summarize the main points and results of the $SU_L(2)$  model of multifermion
couplings and its scaling region proposed in the ref.\cite{xue97,xue99}.
Note that $\psi^i_L$ ($i=1,2$) is an $SU_L(2)$ gauged
doublet, $\chi_R$ is an $SU_L(2)$ singlet and both are two-component Weyl
fermions. $\chi_R$ is treated as a ``spectator'' fermion. 
$\psi^i_L$ and $\chi_R$ fields are dimensionful $[a^{{1\over2}}]$. 
We suggested the following action for the chiral fermions $\psi^i_L$ and $\chi_R$
with the $SU_L(2)\otimes U_R(1)$ chiral symmetries on the lattice:
\begin{eqnarray}
S&=&S_f+S_1+S_2,\label{action}\\
S_f&=&{1\over 2a}\sum_x\sum_\mu\left(\bar\psi^i_L(x)\gamma_\mu D^\mu_{ij}\psi^j_L(x)+
\bar\chi_R(x)\gamma_\mu\partial^\mu\chi_R(x)\right),\nonumber\\
S_1&=&g_1
\sum_x\bar\psi^i_L(x)\cdot\chi_R(x)\bar\chi_R(x)\cdot\psi_L^i(x),
\nonumber\\
S_2&=&g_2\sum_x \bar\psi^i_L(x)\cdot\left[\Delta\chi_R(x)\right]
\left[\Delta\bar\chi_R(x)\right]\cdot\psi_L^i(x),\nonumber
\end{eqnarray}
where $S_f$ is the naive lattice action for chiral fermions, 
$a$ is the lattice spacing fixed as a constant.
$S_1$ and $S_2$ are two external multifermion 
couplings, where the $g_1$ and $g_2$ have dimension
$[a^{-2}]$, and the Wilson factor\cite{wilson} is given as,
\begin{eqnarray}
\Delta\chi_R(x)&\equiv&\sum_\mu
\left[ \chi_R(x+\mu)+\chi_R(x-\mu)-2\chi_R(x)\right],\nonumber\\
2w(p)&=&\int_xe^{-ipx}\Delta(x)=\sum_\mu\left(1-\cos(p_\mu)\right).
\label{wisf}
\end{eqnarray}
Note that all momenta are scaled to be
dimensionless, $p=\tilde p+\pi_A$
where $\pi_A$ runs over fifteen lattice momenta ($\pi_A\not=0$). 

The action (\ref{action}) has an exact local $SU_L(2)$ chiral gauge symmetry,
\begin{equation} 
\sum_\mu \gamma_\mu D^\mu=\sum_\mu(U_\mu(x)\delta_{x,x+\mu}
-U^\dagger_\mu (x)\delta_{x,x-\mu}),\hskip0.5cm U_\mu(x)\in SU_L(2),
\label{kinetic} 
\end{equation} 
which is the gauge symmetry that the continuum theory
(the target theory) possesses. The global flavour symmetry $U_L(1)\otimes
U_R(1)$ is not explicitly broken in eq.~(\ref{action}).
It has been advocated\cite{xue97,xue99} that there exists a plausible scaling region for chiral fermions in the low-energy limit. This
is a peculiar segment in the phase space of the multifermion couplings $g_1$ and $g_2$, 
\begin{equation}
{\cal A}=\Big[g_1\rightarrow 0, g_2^{c,a}<g_2<g_2^{c,\infty}\Big],\hskip0.3cm
a^2g_2^{c,a}=0.124,\hskip0.3cm
1\ll g_2^{c,\infty}< \infty, 
\label{segment}
\end{equation}
where $g_2^{c,\infty}$  is a finite 
number. The crucial
points and results for this scaling region to exist are briefly described in the following.

In the segment ${\cal A}$ (\ref{segment}), the
action (\ref{action}) possesses a global $\chi_R$-shift-symmetry \cite{gp},
i.e., the action is invariant under the transformation:
$\bar\chi_R(x) \rightarrow \bar\chi_R(x)+\bar\delta,
\chi_R(x) \rightarrow \chi_R(x)+\delta,$
where $\delta$ is independent of space-time.  
The Ward identity corresponding to this
$\chi_R$-shift-symmetry is given as\cite{xue97}
($g_1\rightarrow 0$),
\begin{equation}
{1\over 2a}\gamma_\mu\partial^\mu\chi'_R(x)
+g_2\!\langle\Delta\!\left(\bar\psi^i_L(x)\!\cdot\!
\left[\Delta\chi_R(x)\right]\psi_L^i(x)\right)\rangle-{\delta\Gamma\over\delta\bar
\chi'_R(x)}=0.
\label{w}
\end{equation}
Note that for studying Ward identities associated to
the gauge symmetry and other global symmetries in this segment (\ref{segment}), the ``primed'' fermion field $\chi'_R(x),
\psi'_R(x)$, gauge field $A'_\mu(x)$ and the vacuum functional ``$\Gamma$'' are introduced through the generating functional
approach (see eqs.(6-13) in the previous paper\cite{xue97}).

The important consequences 
of this Ward identity (\ref{w}) in segment ${\cal A}$ are:
\begin{itemize}
\item the low-energy mode ($p\sim 0$) of $\chi_R$ is a free mode and decoupled for its one particle irreducible (1PI) function:
\begin{equation}
\int_x e^{-ipx} {\delta^{(2)}\Gamma\over\delta\chi'_R(x)\delta\bar\chi'_R(0)}
={i\over a}\gamma_\mu\sin(p^\mu);
\label{free}
\end{equation}

\item no hard spontaneous chiral symmetry breaking at the lattice scale ($O(1/ a)$) occurs
\footnote{The soft symmetry 
breaking $\tilde v$ at the electroweak scale for the low-energy modes ($p\sim 0$) is allowed and can be achieved by tuning the multifermion couplings $g_1$ and $g_2$.}, as its the self-energy (1PI) function
$\Sigma^i(p)$ vanishes both at $p=0$ (see eqs.(30) and (31) in ref.\cite{xue97}): 
\begin{equation}
\int_x e^{-ipx} {\delta^{(2)}\Gamma\over\delta\psi'^i_L(x)\delta\bar\chi'_R(0)}
={1\over2}\Sigma^i(p)=0 \hskip0.5cm p=0,
\label{ws2}
\end{equation}
and at $p\not=0$ (see eq.(104) in \cite{xue97}):
\begin{equation}
\Sigma(p)=0\hskip0.5cm p\not=0
\label{ws2'}
\end{equation}
which is demonstrated by the strong coupling expansion in the segment ${\cal A}$.

\end{itemize}

On the other hand, for the strong coupling $g_2\gg 1$ in the segment ${\cal A}$, the following 
three-fermion-states
comprising of the elementary fields $\psi^i_L$ and $\chi_R$ in (\ref{action}) are bound:
\begin{equation}
\Psi_R^i={1\over
2a}(\bar\chi_R\cdot\psi^i_L)\chi_R;\hskip1cm\Psi^n_L={1\over 2a}(\bar\psi_L^i
\cdot\chi_R)\psi_L^i.
\label{bound}
\end{equation}
These fermion bound states are Weyl
fermions with the ``wrong'' chiralities in  contrast with 
the ``right'' chiralities possessed by the elementary fields $\psi^i_L$ and $\chi_R$.
By using the strong-coupling expansion in powers of
$1/ g_2$, we compute the following two-point Green functions with 
insertions of appropriate composite operators, for instance, for the charged sector, 
\begin{eqnarray}
S^{ij}_{LL}(x)&=&
\langle\psi^i_L(0)\bar\psi^j_L(x)\rangle,\hskip1cm 
S^{ij}_{RL}(x)=\langle\psi^i_L(0)\bar\Psi^j_R(x)\rangle,\nonumber\\
S^{ij}_{RR}(x)&=&\langle\Psi^i_R(0)\bar\Psi^j_R(x)\rangle,\hskip1cm
S^{ij}_{LR}(x)=\langle\Psi^i_L(0)\bar\psi^j_R(x)\rangle.
\label{twopoint}
\end{eqnarray}
In the lowest nontrivial order, we obtain following recursion relations 
\cite{xue97}, 
\begin{eqnarray} 
S^{ij}_{LL}(x)&=&{1\over g_2\Delta^2(x)}\left({1\over 2a
}\right)^2\sum^\dagger_\mu S^{ij}_{RL}(x+\mu)\gamma_\mu,\label{re11}\\
S^{ij}_{RL}(x)&=&\left({1\over 2a
}\right)\left({\delta(x)\delta_{ij}\over 2g_2\Delta^2(x)}
+{1\over g_2\Delta^2(x)}\left({1\over 2a
}\right)\sum^\dagger_\mu S^{ij}_{LL}(x+\mu)\gamma_\mu\right),
\label{re21}\\
S^{ij}_{RR}(x)&=&\left({1\over 2a
}\right)^2{1\over g_2\Delta^2(x)}\sum^\dagger_\mu 
\gamma_\mu S^{ij}_{RL}(x+\mu).
\label{re31}
\end{eqnarray}
The Fourier transformations of these recursion
equations for $p\not=0$ and $\Delta^2(p)=4w^2(p)\not=0$ lead to,
\begin{eqnarray}
S^{ij}_{LL}(p)&=&\int_x e^{-ipx}\langle\psi^i_L(0)\bar\psi^j_L(x)\rangle
\simeq P_L{\delta_{ij}{i\over a}\sum_\mu\sin p^\mu\gamma_\mu\over
{1\over a^2}\sum_\mu\sin^2 p_\mu+M^2(p)}P_R,\label{sll21}\\
S^{ij}_{RL}(p)&=&\int_xe^{-ipx}\langle\Psi_R^i(0)\bar\psi_L^j(x)\rangle \simeq P_L{\delta_{ij}M(p)\over
{1\over a^2}\sum_\mu\sin^2 p_\mu+M^2(p)}P_L,\label{slm21}\\
S^{ij}_{RR}(p)&=&\int_x e^{-ipx}\langle\Psi^i_R(0)\bar\Psi^j_R(x)\rangle
\simeq P_R{\delta_{ij}{i\over a}\sum_\mu
\sin p^\mu\gamma_\mu\over
{1\over a^2}\sum_\mu\sin^2 p_\mu+M^2(p)}P_L.\label{smm21}\\
M(p)&=&8ag_2w^2(p).\label{m}
\end{eqnarray}
However, these two-point Green functions (\ref{slm21}-\ref{smm21}) show only structure of poles,
but not residues of these poles, which are related to the renormalization of Green functions with insertions of composite fermion operators $\Psi^i_R(x)$. 

The description of the renormalization of n-point
1PI functions with insertions of composite operators shows that
the renormalized n-point 1PI functions $\Gamma_{ren}^{(n)}$ with single and two
insertions of composite operators are given by\cite{itz}, 
\begin{eqnarray}
\Gamma_{ren}^{(n)}(p_1,q_1,q_2,\cdot\cdot\cdot,q_n)&=&
Z\Gamma_{reg}^{(n)}(p_1,q_1,q_2,\cdot\cdot\cdot,q_n),\nonumber\\
\Gamma_{ren}^{(n)}(p_1,p_2,q_1,q_2,\cdot\cdot\cdot,q_n)&=&
Z^2\Gamma_{reg}^{(n)}(p_1,p_2,q_1,q_2,\cdot\cdot\cdot,q_n),
\label{ren}
\end{eqnarray}
where $\Gamma_{reg}^{(n)}$ are the regularized n-point 1PI functions, $p_1$
and $p_2$ stand for the momenta entering the composite operators. 
The renormalization constant $Z$
is the generalized  ``wave-function renormalization'' or ``form factor'' of composite operators.
It is worthwhile to stress that $Z$ is a finite
positive constant and the wave-function renormalization of composite fields 
is the exactly same as the wave-function renormalization of elementary fields.
In fact, composite particles are indistinguishable
from elementary particles in this case. 

To be consistent with this description of the renormalization of n-point
1PI functions with insertions of composite operators, the two-point Green
functions (\ref{slm21}-\ref{smm21}) should be modified as follow, 
\begin{eqnarray}
S^{ij}_{RR}(p)&=&\int_x e^{-ipx}\langle\Psi^i_R(0)\bar\Psi^j_R(x)\rangle
\simeq\delta_{ij}{Z_R(p){i\over a}\sum_\mu\sin p^\mu\gamma_\mu Z_R(p)\over
{1\over a^2}\sum_\mu\sin^2 p_\mu+M^2(p)}P_L;
\label{rsc11}\\
S^{ij}_{RL}(p)
&=&\int_xe^{-ipx}\langle\Psi_R^i(0)\bar\psi_L^j(x)\rangle \simeq \delta_{ij}Z_R(p){M(p)
\over {1\over a^2}\sin^2p + M^2(p)}P_R.
\label{mixingc}
\end{eqnarray}
These regularized two-point Green functions with one and two insertions of composite fermion operators identify not only their poles, but also the corresponding residues.
The residues $Z_R(p)$ (\ref{rsc11}-\ref{mixingc}), i.e., the generalized
form factors of the composite three-fermion-states 
(\ref{bound}), are given by one-particle irreducible (1PI) functions\cite{xue99},
\begin{equation}
Z_R(p)=\int_xe^{-ipx}
{\delta^{(2)}\Gamma\over\delta\Psi'^i_R(x)\delta\bar\psi'^j_L(0)}\simeq aM(p),
\hskip0.5cm Z_L(p)=\int_xe^{-ipx}
{\delta^{(2)}\Gamma\over\delta\Psi'^n_L(x)\delta\bar\chi'_R(0)}=aM(p).
\label{zlzr}
\end{equation}
The ``primed fields'' $\Psi'^n_L(x)$ and $\Psi'^i_R(x)$ of three-fermion-states
are defined by eqs.(41) and (42) in ref.\cite{xue97} through the generating
functional approach and $Z_R(p)$ is obtained\cite{xue99} by the strong coupling expansion for $p\sim\pi_A\not=0$. In eq.(\ref{zlzr}), $Z_L(p)$ is the renormalization constant for the three-fermion-state $\Psi'^n_L(x)$ in the 
neutral sector, which is exactly resulted from the Ward identity (\ref{w}). These residues $Z_{L,R}(p\sim\pi_A)$ (generalized form factors of three-fermion-states $\Psi'^i_R(x), \Psi'^n_L(x)$) are different positive  (non-zero) constants with respect to each 
doubler ($p\simeq\pi_A$).  

We make a wave-function renormalization of 
three-fermion-states with respect to each doubler $p=\pi_A\not=0$\cite{xue99},
\begin{equation}
\Psi_R^i|_{ren}=Z^{-1}_R(p)\Psi_R^i;\hskip1cm\Psi^n_L|_{ren}=Z^{-1}_L(p)\Psi^n_L.
\label{rbound}
\end{equation}
As a result for $p=\pi_A\not=0$, the neutral $\Psi_n$ and 
charged 
$\Psi_c^i$ Dirac fermions are formed, 
\begin{equation}
\Psi^i_c=(\psi_L^i, \Psi^i_R|_{ren});\hskip1cm\Psi_n=(\Psi_L^n|_{ren}, \chi_R),
\label{di}
\end{equation}
whose propagators are
\begin{eqnarray}
S_c^{ij}(p)=\int_xe^{-ipx}\langle\Psi_c^i(0)\bar\Psi^j_c(x)\rangle &\simeq&
\delta_{ij}{{i\over a}\gamma_\mu \sin (p)^\mu +M(p)
\over {1\over a^2}\sin^2p + M^2(p)};\label{dlp'}\\
S_n(p)=\int_xe^{-ipx}\langle\Psi_n(0)\bar\Psi_n(x)\rangle 
&\simeq&{{i\over a}\gamma_\mu \sin (p)^\mu + M(p) 
\over {1\over a^2}\sin^2p + M^2(p)}.
\label{drp'}
\end{eqnarray}
These show that all doublers $(p=\pi_A)$
are decoupled as very massive Dirac fermions consistently with the 
$SU_L(2)\otimes U_R(1)$ chiral symmetries, 
since the three-fermion-states (\ref{bound}) carry the appropriate quantum
numbers of the chiral groups that accommodate $\psi^i_L$ and
$\chi_R$.

Due to the locality of action (\ref{action}),
all Green functions and 1PI functions must be analytically continuous functions in 
energy-momentum space, provided the dynamics is fixed by given multifermion couplings $g_1$ and $g_2$.
Although two-point Green functions (\ref{sll21}) and (\ref{rsc11}-\ref{mixingc}) are obtained from the strong coupling 
expansion $g_2\gg 1$ for the momentum $p=\pi_A\not=0$, we can make analytical 
continuation of these two-point Green functions from $p=\pi_A$ to $p=0$ in the energy-momentum space, provided no hard spontaneous symmetry
breaking takes place (\ref{ws2},\ref{ws2'}) as the effective multifermion coupling $g_2(p)$ is reduced. Since the 
residues $Z_{L,R}(p)$ positively
vanish ($Z_{L,R}(p)\rightarrow O(p^8)$) in the low-energy limit $p\rightarrow 0$, eq.(\ref{rsc11}) has no simple pole at $p\sim 0$ and mixing (\ref{mixingc}) vanishes as well. For $Z_{L,R}(0)=0$, we are not allowed to make the wave-function renormalization
(\ref{rbound}) with respect to $p\sim 0$ and obtain Dirac fermions (\ref{dlp'},\ref{drp'}) at $p\sim 0$. While, the propagator (\ref{sll21}) of the elementary field $\psi_L^i$  
has a simple pole at $p\sim 0$ behaving as a charged chiral particle $\psi^i_L(x)$
consistently with the $SU_L(2)$ symmetry ($\tilde p^\mu$ is the
dimensionful continuum momentum), 
\begin{equation} 
S^{-1}_{LL}(\tilde p)^{ij}=i\gamma_\mu\tilde p^\mu\tilde Z_2\delta_{ij}P_L; 
\hskip0.5cm
S^{-1}_{RR}(\tilde p)=i\gamma_\mu\tilde p^\mu P_R, 
\label{sf} 
\end{equation}
where $\tilde Z_2$ is the finite wave-function renormalization constant of the
elementary interpolating field $\psi_L^i(x)$\cite{xue97} and $S^{-1}_{RR}(\tilde p)$ (\ref{free}) is the inverse propagator of $\chi_R(x)$ at $\tilde p\sim 0$. All these properties discussed were also obtained by the weak coupling expansion in ref.\cite{xue99}. 
  
The vanishing of $Z_{L,R}(p)$ and eqs.(\ref{rsc11}-
\ref{mixingc}) at $p\sim 0$ indicates that three-fermion-states $\Psi_R^i$ and $\Psi_L^n$ 
dissolve into the virtual states of three individual chiral fermions $\psi_L^i(x)$ and $\chi_R(x)$ with the fixed total momentum $p$ and a continuous energy spectrum. The ``no-go'' theorem is entirely inapplicable to these virtual states. 
We call these virtual states three-fermion-cuts ${\cal C}[\Psi_L^n(x)]$ and ${\cal C}[\Psi_R^i(x)]$\cite{xue97l,xue99}. 
These two virtual
states carry the exactly same quantum numbers and total momentum 
$p$ as that of the corresponding three-fermion-states. Thus, chiral gauge symmetries are preserved in such a 
dissolving phenomenon. The energy-threshold $\epsilon$ of such a dissolving, where the energy-gap between three-fermion-states and their corresponding virtual states goes to zero, must locate at  ($\tilde v \ll {\pi\over a}$ being a soft spontaneous symmetry breaking at the electroweak scale)
\begin{equation} 
\tilde v <\epsilon <  {\pi\over a}, 
\label{threshold} 
\end{equation}
whose value depends on the values of the multifermion couplings $g_1$ and $g_2$ in the scaling region ${\cal A}$ (\ref{segment}).

As results, the spectrum of the
model (\ref{action}) in the scaling region ${\cal A}$ (\ref{segment}) is the following. It consists of 15 copies of
$SU_L(2)$-charged Dirac doublers eq.(\ref{dlp'}) and 15 copies of $S_LU(2)$-neutral 
Dirac doublers eq.(\ref{drp'}). They are very massive and decoupled.
Beside, the low energy spectrum contains the massless normal modes
eqs.(\ref{sf}) for $p=\tilde p\sim 0$. 

Before ending this section, we emphasize that our scenario in the scaling region is resulted from the dynamics of the  special dimension-10 operator $S_2$ in the action (\ref{action}). It is worthwhile to 
mention that we can replace the Wilson operator $\Delta$ in $S_2$ by  $\Delta_\mu$ defined as 
\begin{eqnarray}
\Delta_\mu\chi_R(x)&\equiv&\chi_R(x+\mu)-\chi_R(x-\mu),\nonumber\\
2w_\mu(p)&=&\int_xe^{-ipx}\Delta_\mu(x)=\sin(p_\mu),
\label{wisfmu}
\end{eqnarray}
and the multifermion coupling $S_2$ in the action (\ref{action}) is substituted by
\begin{equation} 
\tilde S_2=g_2\sum_{x,\mu} \bar\psi^i_L(x)\cdot\left[\Delta_\mu\chi_R(x)\right]
\left[\Delta_\mu\bar\chi_R(x)\right]\cdot\psi_L^i(x),
\end{equation}
which is a dimension-8 operator. Although this action $\tilde S_2$ has the
exactly same $SU_L(2)$ chiral gauge symmetry, $SU_L(2)\otimes U_R(1)$  global chiral symmetries and $\chi_R$-shift-symmetry, one can check such a dimension-8 operator does not process the proper properties presented in this section. Another example is the multifermion coupling $S_1$ with dimension-6 operators that suffers from a hard spontaneous symmetry breaking as that in ref.\cite{ep}. These two examples tell us that in principle, chiral gauge symmetries can be preserved by multifermion couplings, but in practice, multifermion couplings do not definitely have desired dynamics, which are: (i) non hard spontaneous symmetry breaking; (ii) the dissolving of three-fermion-state to the three-fermion-cut in the low-energy; (iii) doublers gauge-invariantly decoupled as massive particles and (iv) spectator fermions decoupled as free fermions, to have the scaling region for an asymptotic chiral gauge theories in the continuum limit.

\section{The gauge coupling vertices in the neutral sector}

In the computations by the strong coupling expansion to obtain the fermion spectrum discussed in section 2, the $SU_L(2)$ gauge interaction is neglected as its perturbative nature with respect to the strong multifermion coupling $ g_2\gg 1$.  
In the following two sections, we turn on the gauge field as a dynamical field to examine (i) whether the spectator field $\chi_R$ decouples from the gauge field; (ii) whether charged chiral fermion couples to the gauge field in chiral manner, as required by an asymptotic chiral gauge theory in the continuum limit.

As the $SU_L(2)$ chiral gauge coupling $g$ is 
turned on and the action (\ref{action}) is $SU_L(2)$-chirally gauged, the properties of the scaling region ${\cal A}$ (\ref{segment}) should not be greatly altered for the reasons that the
$SU_L(2)$-chiral gauge interaction does not spoil the Ward identity (\ref{w}) of the $\chi_R$-shift-symmetry and the spectrum is gauge symmetric  in the scaling region ${\cal A}$ (\ref{segment}).

Now we consider all possible interacting vertices (one particle irreducible (1PI) functions) involving truncated external gauge field $A'_\mu$ and fermion fields $\chi'_R, \Psi'^n_L$ in the neutral sector. Based on the Ward identity (\ref{w}) of the $\chi_R$-shift-symmetry, we take
functional derivatives with respect of the external gauge field $A'_\mu$, fermion fields $\chi'_R$ and $\Psi'^n_L$, to obtain the following Ward identities, 
\begin{equation}
{\delta^{(2)} \Gamma\over\delta
A'_\mu\delta\bar\chi'_R}={\delta^{(3)} \Gamma\over\delta
A'_\mu\delta\chi'_R\delta\bar\psi'_R}={\delta^{(3)} \Gamma\over\delta
A'_\mu\delta\Psi'^n_L\delta\bar\chi'_R}=\cdot\cdot\cdot=0.
\label{wa}
\end{equation}
As a consequence of these Ward identities and identical vanishing of interacting 1PI functions
containing external gauge field $A'_\mu$, ``spectator'' fermions $\chi'_R(x)$ and neutral
three-fermion states field $\Psi'^n_L(x)$, we prove no any interactions
between the gauge field $A_\mu$, the ``spectator'' fermion $\chi_R$ and the
neutral three-fermion states $\Psi_L^n(x)$. It should be emphasized that the decoupling of the neutral sector ($\chi_R,\Psi^n_L$) from the gauge field $A_\mu$ is valid not
only for perturbative gauge-interaction but also non-perturbative gauge-interaction, since the action is exactly $SU_L(2)$-chiral-gauge symmetric and $\chi_R$-shift-symmetric for any values of the gauge coupling and multifermion couplings.

As a result, The right-handed current $j^\mu_R=i\bar\chi_R\gamma^\mu\chi_R$ is exactly conserved 
i.e., $\partial_\mu j^\mu =0$ and no anomalous contribution is expected from the topological gauge 
configuration. This is a very important feature, which we will see later in section 5 and 6, for obtaining the gauge anomaly and the fermion-flavour singlet anomaly relating to the $U_L(1)$-symmetry (the number of the fermion field $\psi_L^i$) violation.    

\section{Gauge coupling vertices in the charged sector}

In this section we turn to directly compute three-point interacting vertices of gauge field $A_\nu^a(x)$, elementary and composite fermion fields $\psi^i_L(x)$ and $\Psi^i_R(x)$, in order to show the gauge-fermion coupling is chiral in the low-energy region, while vector-like in the high-energy region. 

Setting 1PI vertex functions of the gauge-fermion coupling to be $\Lambda_\mu^a(p,p')$ and $q=p'+p$, where $p'$ and $p$ are two external momenta of fermion fields and $q$ is the momentum of the gauge field, 
we can write the following three-point Green functions in the momentum space:
\begin{eqnarray}
\int_{x_1xy}e^{i(p'x\!+\!px_1\!-\!qy)}\langle\psi_L(x_1)\bar\psi_L(x) A_\nu^a(y)\rangle
\!&=&\! G^{ab}_{\nu\mu}(q)S_{LL}(p) \Lambda^b_{\mu LL}(p,p')S_{LL}(p');
\label{rpa1}\\
\int_{x_1xy}e^{i(p'x\!+\!px_1\!-\!qy)}\langle\psi_L(x_1)\bar\Psi_R(x) A_\nu^a(y)\rangle
\!&=&\! G^{ab}_{\nu\mu}(q)S_{LL}(p) \Lambda^b_{\mu LR}(p,p')S_{RR}(p');
\label{rpa2}\\
\int_{x_1xy}e^{i(p'x\!+\!px_1\!-\!qy)}\langle\Psi_R(x_1)\bar\Psi_L(x) A_\nu^a(y)\rangle
\!&=&\! G^{ab}_{\nu\mu}(q)S_{RR}(p) \Lambda^b_{\mu RL}(p,p')S_{LL}(p');
\label{rpa4}\\
\int_{x_1xy}e^{i(p'x\!+\!px_1\!-\!qy)}\langle\Psi_R(x_1)\bar\Psi_R(x) A_\nu^a(y)\rangle
\!&=&\! G^{ab}_{\nu\mu}(q)S_{RR}(p) \Lambda^b_{\mu RR}(p,p')S_{RR}(p'),
\label{rpa3}
\end{eqnarray}
where $G^{ab}_{\nu\mu}(q)$ is the propagator of  gauge field $A_\nu^a(x)$; the 
$S_{LL}(p)$ and $S_{RR}(p)$ are the propagators of the elementary and composite chiral fermions 
$\psi_L(x)$ and $\Psi_R(x)$, we omit henceforth the $SU_L(2)$ indices $i$ and $j$. We try to compute the 1PI
vertex functions $\Lambda^b_{\mu LL}(p,p'), \Lambda^b_{\mu RR}(p,p')$, $\Lambda^b_{\mu LR}(p,p')$ and $\Lambda^b_{\mu RL}(p,p')$ in eqs.(\ref{rpa1}-\ref{rpa3}).

Using the small gauge coupling ($g$) expansion, one can find the 1PI
vertex functions $\Lambda^b_{\mu LL}(p,p')$ by calculating
\begin{eqnarray}
&&\langle\psi_L(x_1)\bar\psi_L(x) A^a_\mu(y)\rangle=i{g\over 2}\left(
{\tau^a\over 2}\right)\langle\psi_L(x_1)\bar\psi_L(x)\rangle\gamma_
\rho\nonumber\\
&&\int_z\left[\langle\psi_L(z\!+\!\rho)\bar\psi_L(x)\rangle\langle
A^b_\rho(z\!+\!{\rho\over2})A^a_\mu(y)\rangle\!+\!
\langle\psi_L(z\!-\!\rho)\bar\psi_L(x)\rangle\langle
A^b_\rho(z\!-\!{\rho\over2})A^a_\mu(y)\rangle\right],
\label{3p}
\end{eqnarray}
and obtains 
\begin{eqnarray}
\Lambda_{\mu LL}^{(1)}(p,p') &=& ig\left(\tau^a\over 2\right)\cos\left({p+p'\over 2}
\right)_\mu\gamma_\mu P_L,\label{normal}\\
\Lambda_{\mu\nu LL}^{(2)}(p,p') &=& -i{g^2\over 2}\left(\tau^a\tau^b\over 4\right)
\sin\left({p+p'\over 2}\right)_\mu
\gamma_\mu \delta_{\mu\nu}P_L,\nonumber\\
&\cdot\cdot\cdot,&\nonumber
\end{eqnarray}
in terms of the powers of the small gauge coupling $(g)$.

Then, we try to find the relationships between the 1PI
vertex function $\Lambda^b_{\mu LL}(p,p')$ and other three  1PI
vertex functions $\Lambda^b_{\mu RR}(p,p'),\Lambda^b_{\mu LR}(p,p'),\Lambda^b_{\mu RL}(p,p')$ in eqs.(\ref{rpa2}-\ref{rpa3}).
By the strong coupling expansion in powers of $1/ g_2$ for $p,p'\sim\pi_A\not=0$, analogously to recursion
relations (\ref{re11}-\ref{re31}), we obtain the following recursion relations
at the nontrivial order, 
\begin{eqnarray}
\langle\psi_L(x_1)\bar\psi_L(x) A_\nu^a(y)\rangle
&=&{1\over g_2\Delta^2(x)}\left({1\over 2a}\right)^2\sum^\dagger_\rho
\langle\psi_L(x_1)\bar\Psi_R(x+\rho) A_\nu^a(y)\rangle\gamma_\rho,
\label{ra1}\\
\langle\psi_L(x_1)\bar\psi_L(x) A_\nu^a(y)\rangle
&=&{1\over g_2\Delta^2(x_1)}\left({1\over 2a}\right)^2\sum^\dagger_\rho
\gamma_\rho\langle\Psi_R(x_1+\rho)\bar\psi_L(x) A_\nu^a(y)\rangle,
\label{ra1'}\\
\langle\Psi_R(x_1)\bar\Psi_R(x) A_\nu^a(y)\rangle
&=&{1\over g_2\Delta^2(x)}\left({1\over 2a}\right)^2\sum^\dagger_\rho\gamma_\rho
\langle\psi_L(x_1)\bar\Psi_R(x+\rho) A_\nu^a(y)\rangle.
\label{ra3}
\end{eqnarray}
We make the Fourier transform in both sides of the above recursion relations and obtain for $p,p'\sim\pi_A\not=0$,  
\begin{eqnarray}
S_{LL}(p) \Lambda^a_{\mu LL}(p,p')S_{LL}(p')
&=&{i\over aM(p')}
S_{LL}(p) \Lambda^a_{\mu LR}(p,p')S_{RR}(p')
\sum_\rho\sin p'_\rho\gamma^\rho,
\label{ra1p}\\
S_{LL}(p) \Lambda^a_{\mu LL}(p,p')S_{LL}(p')
&=&{i\over aM(p)}\sum_\rho\sin p_\rho\gamma^\rho
S_{RR}(p) \Lambda^a_{\mu RL}(p,p')S_{LL}(p'),
\label{ra1p'}\\
S_{RR}(p) \Lambda^a_{\mu RR}(p,p')S_{RR}(p')
&=&{i\over aM(p')}
\sum_\rho\sin p'_\rho\gamma^\rho S_{LL}(p)\Lambda^a_{\mu LR}(p,p')S_{RR}(p'),
\label{ra3p}
\end{eqnarray}
where the propagator of the gauge boson $G^{ab}_{\nu\mu}(q)$ is
eliminated from the both sides of equations. 

Using these recursion relations (\ref{ra1p}-\ref{ra3p}) and the propagators $S_{LL}(p),S_{RR}(p)$ (\ref{sll21},\ref{smm21}) obtained by the same strong coupling expansion $1/g_2$ for $p\sim\pi_A\not=0$, we can compute the 1PI
vertex functions $\Lambda^a_{\mu RL}(p,p')$, $\Lambda^a_{\mu LR}(p,p')$ and
$\Lambda^a_{\mu RR}(p,p')$ eqs.(\ref{rpa2}-\ref{rpa3}) in terms of the 1PI vertex function
$\Lambda^a_{\mu LL}(p,p')$),
\begin{eqnarray}
M(p')\Lambda^a_{\mu LL}(p,p')
&=&\Lambda^a_{\mu LR}(p,p')\left({i\over a}\right)
\sum_\rho\sin p'_\rho\gamma^\rho,
\label{v1p}\\
M(p)\Lambda^a_{\mu LL}(p,p')
&=&\left({i\over a}\right)\sum_\rho\sin p_\rho\gamma^\rho
\Lambda^a_{\mu RL}(p,p'),
\label{v1p'}\\
M(p')\Lambda^a_{\mu RR}(p,p')
&=&\left({i\over a}\right)
\sum_\rho\sin p'_\rho\gamma^\rho \Lambda^a_{\mu LR}(p,p').
\label{v3p}
\end{eqnarray}
These relationships are independent of the strength of the gauge coupling $g$.
Taking $\Lambda^a_{\mu LL}(p,p')$ to be eq.(\ref{normal}) at the leading 
order, we obtain perturbative results
\begin{eqnarray}
\Lambda_{\mu RR}^{(1)}(p,p') &=& ig\left(\tau^a\over 2\right)\cos\left({p+p'\over 2}
\right)_\mu\gamma_\mu P_R,\label{vr}\\
\Lambda_{\mu LR}^{(1)}(p,p')\left({i\over a}\right)
\sin p'_\mu &=& {1\over2}M(p')ig\left(\tau^a\over 2\right)
\cos\left({p+p'\over 2}
\right)_\mu ,\label{vlr}\\
\left({i\over a}\right)
\sin p_\mu\Lambda_{\mu RL}^{(1)}(p,p') &=& {1\over2}M(p)ig\left(\tau^a\over 2\right)
\cos\left({p+p'\over 2}
\right)_\mu.
\label{vrl}
\end{eqnarray}

According to the renormalization of 1PI functions with composite operator insertions (\ref{ren}), we
have the renormalized 1PI vertex functions for $p,p'\not=0$ (see Figure 5 in Ref.\cite{xue99}), 
\begin{eqnarray}
\Lambda^a_{\mu RR}(p,p')|_{ren}
&=&Z_R(p)Z_R(p')\Lambda^a_{\mu RR}(p,p'),
\label{rv1p}\\
\Lambda^a_{\mu RL}(p,p')|_{ren}
&=&Z_R(p)\Lambda^a_{\mu RL}(p,p'), 
\label{rv1p''}\\
\Lambda^a_{\mu LR}(p,p')|_{ren}
&=&Z_R(p')\Lambda^a_{\mu LR}(p,p'), 
\label{rv1p'}
\end{eqnarray}
where $\Lambda^a_{\mu RR}(p,p'), \Lambda^a_{\mu LR}(p,p')$ and $\Lambda^a_{\mu RL}(p,p')$ are the regularized 1PI vertex functions obtained in eqs.(\ref{v1p}-\ref{v3p}) in terms of $\Lambda_{\mu LL}(p,p')$. For $p,p'=\pi_A\not=0$ with respect to each doublers, $Z_R(p)$ and $Z_R(p')\not=0$, and we can make the renormalization (\ref{rbound}) of the composite fermion-operator (field) $\Psi_R(x)$ so that,
\begin{eqnarray}
\Lambda^a_{\mu RR}(p,p',\Psi_R|_{ren})
&=&\Lambda^a_{\mu RR}(p,p'),
\label{rrv1p}\\
\Lambda^a_{\mu RL}(p,p',\Psi_R|_{ren})
&=&\Lambda^a_{\mu RL}(p,p'), 
\label{rrv1p''}\\
\Lambda^a_{\mu LR}(p,p',\Psi_R|_{ren})
&=&\Lambda^a_{\mu LR}(p,p'), 
\label{rrv1p'}
\end{eqnarray}
where $\Lambda^a_{\mu RR}(p,p',\Psi_R|_{ren}), \Lambda^a_{\mu LR}(p,p',\Psi_R|_{ren})$ and $\Lambda^a_{\mu RL}(p,p',\Psi_R|_{ren})$ are the renormalized 1PI vertex functions in terms of the renormalized composite fermion-operator(field) $\Psi_R|_{ren}$ (\ref{rbound}). Correspondingly, the charged 
Dirac fermion $\Psi_c=(\psi_L, \Psi_R|_{ren})$ and its propagator $S_c(p)$ in terms of the renormalized composite fermion-operator(field) $\Psi_R|_{ren}$ (\ref{rbound}) is given by eq.(\ref{dlp'}) in section 2.

The interacting 1PI vertex function between the gauge 
field $A_\mu^a(x)$ and the charged Dirac fermion $\Psi_c(x)$ is related to the 
following three-point Green functions,
\begin{eqnarray}
\langle\Psi_c(x_1)\bar\Psi_c(x) A_\nu^a(y)\rangle
&=&\langle\psi_L(x_1)\bar\psi_L(x) A_\nu^a(y)\rangle
+\langle\psi_L(x_1)\bar\Psi_R(x)|_{ren} A_\nu^a(y)\rangle
\nonumber\\
&+&\langle\Psi_R(x_1)|_{ren}\bar\psi_L(x) A_\nu^a(y)\rangle
+\langle\Psi_R(x_1)|_{ren}\bar\Psi_R(x)|_{ren} A_\nu^a(y)\rangle,
\label{3points}
\end{eqnarray}
and
\begin{equation} 
\int_{x_1xy}e^{i(p'x\!+\!px_1\!-\!qy)}\langle\Psi_c(x_1)\bar\Psi_c(x) A_\nu^a(y)\rangle
\!=\! G^{ab}_{\nu\mu}(q)S_c(p) \Lambda^b_{\mu c}(p,p',\Psi_R|_{ren})S_c(p'),
\label{rpad}
\end{equation}
where $\Lambda^b_{\mu c}(p,p',\Psi_R|_{ren})$ is the renormalized 1PI vertex function of the gauge field $A_\mu^a(x)$ and the charged Dirac fermion $\Psi_c=(\psi_L, \Psi_R|_{ren})$. 

From eqs.(\ref{rpa1}-\ref{rpa3}) and eqs.(\ref{3points}-\ref{rpad}), the 1PI vertex function $\Lambda_{\mu c}(p,p',\Psi_R|_{ren})$ is given by
\begin{eqnarray}
\Lambda_{\mu c}(p,p',\Psi_R|_{ren})^{(1)}&=&\Lambda_{\mu LL}(p,p')^{(1)}+\Lambda_{\mu LR}(p,p',\Psi_R|_{ren})^{(1)}
+\Lambda_{\mu RL}(p,p',\Psi_R|_{ren})^{(1)}\nonumber\\
&+&\Lambda_{\mu RR}(p,p',\Psi_R|_{ren})^{(1)},
\label{vdirac}
\end{eqnarray}
up to the first order of the perturbative gauge coupling $g$. One can check that the 1PI vertex functions (\ref{vr}-\ref{vrl}) and the renormalization for these vertex function (\ref{rv1p}-\ref{rv1p'}) with respect to each doublers $p',p=\pi_A\not= 0$ precisely obey the following extremely important Ward identity of
the exact $SU_L(2)$ chiral gauge symmetry at the lattice scale $\pi/ a$,
\begin{equation}
\left({i\over a}\right)(
\sin p_\mu-\sin p'_\mu)\Lambda_{\mu c}^{(1)}(p,p',\Psi_R|_{ren})=S_c^{-1}(p)-S_c^{-1}(p').
\label{gward}
\end{equation}
where the gauge coupling $g$ and gauge group generator $\tau_a/ 2$ are eliminated
from the vertex function $\Lambda_{\mu c}(p,p',\Psi_R|_{ren})$. This shows that the exact $SU_L(2)$ chiral gauge symmetry is
realized by the vector-like and massive spectrum of Dirac modes $p',p=\pi_A\not= 0$ (\ref{di}).
Such results are expected, since we are in the symmetric phase
($1\ll g_2<\infty$). These calculations can be straightforwardly generalized to higher orders of
the perturbative expansion in powers of the gauge coupling $g$. 

So far, by the strong coupling expansion for $p',p=\pi_A\not= 0$, we have computed the 1PI regularized vertex functions $\Lambda^b_{\mu RR}(p,p'), \Lambda^b_{\mu LR}(p,p')$ and $\Lambda^b_{\mu RL}(p,p')$ (\ref{v1p}-\ref{v3p}) in terms of $\Lambda^b_{\mu LL}(p,p')$ (\ref{rpa1},\ref{normal}), and the corresponding 1PI renormalized vertex functions (\ref{rv1p})-(\ref{rv1p'}). Clearly the computations by the strong coupling expansion are broken down for $p', p\rightarrow 0$. However, these 1PI renormalized vertex functions (\ref{rrv1p}-\ref{rrv1p'}) are analytical continuous functions in the energy-momentum plane for the locality of the action (\ref{action}). We can make an analytical continuation of the 1PI renormalized vertex functions (\ref{rrv1p})-(\ref{rrv1p'}) from $p',p=\pi_A\not= 0$ to $p',p\sim 0$, provided no hard spontaneous symmetry breaking takes place (\ref{ws2},\ref{ws2'}). Eqs.(\ref{rv1p}-\ref{rv1p'}) show that the renormalized 1PI vertex functions 
$\Lambda^a_{\mu RR}(p,p')|_{ren}, \Lambda^a_{\mu LR}(p,p')|_{ren}$ and $\Lambda^a_{\mu RL}(p,p')|_{ren}$  vanish as $Z_R(p)$ and $Z_R(p')$ vanish for $p', p\rightarrow 0$.
Furthermore, due to $Z_R(p)$ vanishes for $p\rightarrow 0$, we are not allowed to make the renormalization of the composite fermion-operator(field) $\Psi_R(x)$ (\ref{rbound}) to obtain eqs.(\ref{rrv1p}-\ref{rrv1p'}) and (\ref{vdirac}). As a result, the interacting 1PI vertex function between the gauge field $A_\nu^a(x)$ and charged
fermion turns out to be $\Lambda_{\mu LL}(p,p')$ for $p', p\rightarrow 0$ and the Ward identity
(\ref{gward}) is reduced to its counterpart of the continuum theory,  
\begin{equation}
\left({i\over a}\right)(
\sin p_\mu-\sin p'_\mu)\Lambda_{\mu LL}^{(1)}(p,p')
=S_{LL}^{-1}(p)-S_{LL}^{-1}(p'),\hskip0.5cm p', p\rightarrow 0
\label{glward}
\end{equation}
where the propagator $S_{LL}(p)$ of elementary chiral fermion $\psi_L^i(x)$ is given by
eq.(\ref{sll21}). We recall again that two-point functions (\ref{rsc11}) and (\ref{mixingc}) vanish as $p$  goes to zero. This Ward identity (\ref{glward}) is consistent with the $SU_L(2)$ chiral gauge
symmetry realized by the chiral spectrum. 

The dissolving of the composite three-fermion-state $\Psi_R(x)$ into its three-fermion-cut ${\cal C}[\Psi_R(x)]$ at the energy-threshold (\ref{threshold}) described in the previous section, results in the vanishing of the 1PI vertex functions $\Lambda^a_{\mu RR}(p,p')|_{ren}, \Lambda^a_{\mu LR}(p,p')|_{ren}$ and $\Lambda^a_{\mu RL}(p,p')|_{ren}$in eqs.(\ref{rv1p}-\ref{rv1p'}). This implies the decoupling between the gauge field $A_\mu^a(x)$ and the composite three-fermion-state $\Psi_R(x)$ (\ref{bound}) in the low-energy. 

All discussions in this section are on the basis of (i) the 1PI vertex functions (\ref{rv1p},\ref{rv1p'}) computed by the strong coupling expansion at $p',p=\pi_A\not= 0$ and (ii) the 1PI vertex functions (\ref{rv1p}-\ref{rv1p'}) analytically extrapolated to $p',p\sim 0$ by the analytical continuation property of these 1PI vertex functions in the energy-momentum plane in the segment ${\cal A}$ (\ref{segment}). However, it is worthwhile on the other hand to directly compute these 1PI vertex functions in the region of $p', p\rightarrow 0$ by the weak coupling expansion. 

By the weak coupling expansion for $p', p\rightarrow 0$, we can directly compute these 1PI vertex functions (\ref{rv1p},\ref{rv1p'}) to see whether the consistent results can be reached. For the small gauge coupling, analogous to eq.(\ref{3p}), the three-point Green functions with the insertions of composite three-fermion-operator(field) are given by,
\begin{eqnarray}
&&\langle\Psi_R(x_1)\bar\Psi_R(x) A^a_\mu(y)\rangle=i{g\over 2}\left(
{\tau^a\over 2}\right)\langle\Psi_R(x_1)\bar\Psi_R(x)\rangle\gamma_
\rho\nonumber\\
&&\int_z\left[\langle\Psi_R(z\!+\!\rho)\bar\Psi_R(x)\rangle\langle
A^b_\rho(z\!+\!{\rho\over2})A^a_\mu(y)\rangle\!+\!
\langle\Psi_R(z\!-\!\rho)\bar\Psi_R(x)\rangle\langle
A^b_\rho(z\!-\!{\rho\over2})A^a_\mu(y)\rangle\right],
\label{3p3}
\end{eqnarray}
and
\begin{eqnarray}
&&\langle\psi_L(x_1)\bar\Psi_R(x) A^a_\mu(y)\rangle=i{g\over 2}\left(
{\tau^a\over 2}\right)\langle\psi_L(x_1)\bar\Psi_R(x)\rangle\gamma_
\rho\nonumber\\
&&\int_z\left[\langle\psi_L(z\!+\!\rho)\bar\Psi_R(x)\rangle\langle
A^b_\rho(z\!+\!{\rho\over2})A^a_\mu(y)\rangle\!+\!
\langle\psi_L(z\!-\!\rho)\bar\Psi_R(x)\rangle\langle
A^b_\rho(z\!-\!{\rho\over2})A^a_\mu(y)\rangle\right].
\label{3p3'}
\end{eqnarray}
Actually, in the section 6 of ref.\cite{xue99}, we made the calculations of two-point Green functions $\langle\Psi_R(x_1)\bar\Psi_R(x)\rangle$ and $\langle\psi_L(x_1)\bar\Psi_R(x)\rangle$ by the weak coupling expansion for the small momenta $p\sim 0$ in the segment ${\cal A}$ (\ref{segment}). The results show that these two-point Green functions in the momentum space,
\begin{equation}
S_{RR}(p)=\int_x e^{-ipx}\langle\Psi_R(0)\bar\Psi_R(x)\rangle,\hskip0.3cm
S_{LR}(p)=\int_x e^{-ipx}\langle\psi_L(0)\bar\Psi_R(x)\rangle
\label{rrcut}
\end{equation}
have no simple poles at $p\sim 0$ contrasting with that in eq.(\ref{slm21},\ref{smm21}), instead are regular and vanish as $p\rightarrow 0$. As a direct consequence, the 1PI vertex functions $\Lambda^a_{\mu RR}(p,p')|_{ren}$ and $ \Lambda^a_{\mu LR}(p,p')|_{ren}$ relating to the three-point Green functions (\ref{3p3},\ref{3p3'}) vanish as their external momenta $p$ and $p'$ vanish. The Ward identity (\ref{gward}) is reduced to (\ref{glward}) for the small external momenta $p,p'\sim 0$. 
The results are the same as that obtained by the strong coupling expansion and the analytical continuation of these 1PI vertex functions in the momentum space.

In conclusion, based on the computations and discussions of the spectrum  and 1PI gauge-coupling vertices of the model (\ref{action}) in the segment ${\cal A}$ (\ref{segment}), we demonstrate the following scenario:
\begin{itemize} 
\item
fifteen non-degenerate massive Dirac fermions (\ref{dlp'}) in the high-energy $p\simeq\pi_A\not= 0$ vectorially coupling to the gauge field consistently with the $SU_L(2)$ chiral gauge
symmetry;  
\item
one massless chiral fermion $\psi_L^i$ (\ref{dlp'}) in the low-energy $p\simeq 0$ chirally coupling to the $SU_L(2)$ gauge field;  
\item 
the neutral fermion sector of
Dirac doublers (\ref{drp'}) and the spectator $\chi_R$ entirely decoupling from the gauge field. 
\end{itemize}
The $SU_L(2)$ chiral gauge
symmetry and $SU_L(2)\otimes U_R(1)$ global chiral symmetry are exactly preserved at the lattice scale for not only perturbative but also non-perturbative gauge field configurations. It is no doubt that we need to have numerical
simulations to show if such a scenario is indeed realized in the scaling region
${\cal A}$ of the action (\ref{action}) proposed.   
 
\section{The vacuum functional and gauge anomaly}

The Ward identities (\ref{gward}) and (\ref{glward}) play an extremely
important r\^ole in a guarantee that the gauge perturbation theory in the scaling
region ${\cal A}$ (\ref{segment}) is gauge symmetric. To all orders of the gauge coupling
perturbation theory, gauge boson masses vanish and the gauge boson propagator
is gauge-invariantly transverse. The gauge perturbation theory can be described
by the normal renormalization prescription as that of normal vector-like gauge 
theories, as well as
that of gauge theories with the soft spontaneous symmetry breaking like the Standard Model in the 
continuum.
In fact, due to the manifest $SU_L(2)$ chiral gauge symmetry and corresponding
Ward identities that are respected by the spectra (vector-spectrum for
$p\sim\pi_A\not=0$ and chiral-spectrum for $p\simeq 0$) in this scaling region, we
should then apply the Rome approach \cite{rome} (which is based on the
conventional wisdom of quantum field theories) to the perturbation theory in the
small gauge coupling. It is expected that the Rome approach would work in the
same way but all gauge-variant (hard gauge-symmetry-breaking) counterterms are 
prohibited.

Provided the scenario of the gauge coupling vertex and spectrum given in above
sections, we find that the gauge field not only chirally couples
to the massless chiral fermion of the $\psi_L^i$ in the low-energy region, but
also vectorially couples to the massive doublers of Dirac fermion $\Psi_c^i$ in
the high-energy regime. In this section, we perturbatively compute the vacuum functional in the small gauge coupling ($g$), and discuss the gauge anomaly and the
renormalization prescription of the gauge-symmetric perturbation theory in this scenario. 

We consider the following $n$-point 1PI functional:
\begin{equation}
\Gamma^{(n)}_{\{\mu\}}=
{\delta^{(n)}\Gamma(A')\over\delta A'_{\mu_1}(x_1)\cdot\cdot\cdot\delta 
A'_{\mu_j}
(x_j)\cdot\cdot\cdot\delta A'_{\mu_n}(x_n)},
\label{fun}
\end{equation}
where $j=1\cdot\cdot\cdot n, (n\geq 2)$ and $\Gamma(A')$ is the vacuum
functional of the external gauge field $A'$. The perturbative computations of the 1PI vertex functions
$\Gamma^{(n)}_{\{\mu\}}$ can be straightforwardly performed by adopting the
method presented in ref.\cite{smit82} for the lattice QCD. Dividing the integration
of internal momenta (internal fermion loop) into 16 hypercubes where 16 modes live, we have 16
contributions to the truncated n-point 1PI functional. The region for the chiral
fermion modes of continuum limit is defined as
\begin{equation}
\Omega=[0,\epsilon]^4,\hskip0.5cm p<\epsilon\ll {\pi\over 2},\hskip0.5cm
p\rightarrow 0, 
\label{continuum}
\end{equation}
where the $\epsilon$ is the energy-threshold (\ref{threshold}) where the three-fermion-state $\Psi_R(x)$ dissolve to the three-fermion-cut ${\cal C}[\Psi_R(x)]$.

As a first example, we deal with the vacuum polarization (in the following we refer $p$ to the external momentum of gauge bosons),
\begin{equation}
\Pi_{\mu\nu}(p)=\sum_{i=1}^{16}\Pi^i_{\mu\nu}(p)=\Pi^c_{\mu\nu}(p)+\Pi^d_{\mu\nu}(p),\hskip0.5cm
\Pi^d_{\mu\nu}(p)=\sum_{i=2}^{16}\Pi^i_{\mu\nu}(p),
\label{vacuum}
\end{equation}
where $\Pi^d_{\mu\nu}(p)$ is doublers' contributions and $\Pi^c_{\mu\nu}(p)$ the contribution from the massless chiral mode in the region (\ref{continuum}). As
for the contributions
$\Pi^d_{\mu\nu}(p)$ from the 15 doublers $(i=2,...,16)$, we make a Taylor
expansion in terms of external physical momenta $p=\tilde p$ and the following
equation is {\it mutatis mutandis} valid 
\cite{smit82},
\begin{eqnarray}
\Pi^d_{\mu\nu}(p)&=&\Pi^\circ_{\mu\nu}(0)+\Pi^{d(2)}_{\mu\nu}(p)(\delta_{\mu\nu}
p^2-p_\mu p_\nu)\nonumber\\
&+&\sum^{16}_{i=2}\left(1-p_\rho|_\circ\partial_\rho-{1\over2}
 p_\rho p_\sigma|_\circ\partial_\rho\partial_\sigma\right)
\Pi^{con}_{\mu\nu}(p,m_i),
\label{smit}
\end{eqnarray}
where $|_\circ f(p)=f(0)$ and $m^i$ are doubler masses. The first and second terms are specific for the
lattice regularization. Since the 15 doublers are gauged as an $SU_L(2)$ vector-like
gauge theory with propagator (\ref{dlp'}) and interacting vertex (\ref{vdirac}), the
Ward identity (\ref{gward}) associated with this vectorlike gauge symmetry results in the
vanishing of the first divergent term $\Pi^\circ_{\mu\nu}(0)$ and the gauge
invariance of the second finite contact term in eq.(\ref{smit}). We recall
that in Roma approach, this was achieved by adding gauge-variant counterterms at the lattice scale to enforce the valid of Ward identities. The third term
in eq.(\ref{smit}) corresponds to the relativistic contribution of the 15
doublers. The $\Pi^{con}_{\mu\nu}(p,m_i)$ is logarithmically divergent and
evaluated in some continuum regularization for vectorlike gauge theories. For doubler masses $m_i$ of
$O(a^{-1})$, the third term in eq.(\ref{smit}) is just finite and
gauge-invariant contributions. 

We turn to the contribution $\Pi^c_{\mu\nu}(p)$ from the massless
chiral mode that is in the first hypercube $\Omega=[-\epsilon,\epsilon]^4$
(\ref{continuum}).
We can use some continuum regularization to calculate this 
contribution, 
\begin{equation}
\Pi^c_{\mu\nu}(p)=\Pi^{c(2)}_{\mu\nu}(p)(\delta_{\mu\nu}
p^2-p_\mu p_\nu),
\label{nsmit}
\end{equation}
up to some finite local counterterms that are subtracted away in the normal renormalization prescription. The spectrum eq.(\ref{sf}) and gauge-coupling vertex eq.(\ref{normal}) with 
respect to
the chiral mode are $SU_L(2)$ chiral-gauge symmetric. The Ward identity (\ref{glward})
associated with this chiral gauge symmetry makes eq.(\ref{nsmit}) to be gauge
invariant and the gauge boson mass is zero to all orders of perturbative calculations. The $\epsilon$-dependence (logarithmical divergence $\ln\epsilon$) in eq.(\ref{nsmit}) has to be exactly
canceled
out from those contributions (\ref{smit}) from doublers, because 
the continuity of 1PI vertex functions in the momentum 
space. 
In summary, the total vacuum polarization $\Pi_{\mu\nu}(p)$ contains two parts:
(i) the vacuum polarization of the chiral mode $\psi^i_L$ in some continuum 
regularization; (ii) gauge invariant finite terms stemming from doublers'
contributions. The second part is the same as the perturbative lattice 
QCD, and can be subtracted away in the normal renormalization prescription.

The second example is the 1PI vertex functions $\Gamma^{(n)}_{\{\mu\}}(\{p\}) (n\geq
4)$,
\begin{eqnarray}
\Gamma^{(n)}_{\{\mu\}}(\{p\})&=&\sum^{16}_{i=1}
\Gamma^{(n)i}_{\{\mu\}}(\{p\},m_i)\hskip0.5cm n\geq 4,
\label{n=4}\\
\{p\}&=&p_1,p_2,\cdot\cdot\cdot\nonumber\\
\{\mu\}&=&\mu_1,\mu_2,\cdot\cdot\cdot,\nonumber
\end{eqnarray}
where the internal momentum integral is analogously divided into the contributions
from sixteen sub-regions of the Brillouin zone where sixteen modes live. Based on the gauge invariance
and power counting, one concludes that up to some gauge invariant finite terms,
the $\Gamma^{(n)}_{\{\mu\}}(\{p\}) (n\geq 4)$ (\ref{n=4}) contain the 15
continuum expressions for 15 massive ($m_i$) Dirac doublers and one for the
massless Weyl mode. The 15 doubler contributions vanish for $m_i\sim
O(a^{-1})$. The $n$-point 1PI vertex functions (\ref{n=4}) end up with their
continuum counterpart for the Weyl fermion and some gauge invariant finite
terms. These finite gauge invariant terms come from  doublers' contributions
are similar to those in the lattice QCD, and can be subtracted away in the
normal renormalization prescription. 

The most important contribution to the vacuum functional is the triangle graph
$\Gamma_{\mu\nu\alpha} (p,q)$. Again, dividing the
integration of the internal momenta into 16 hypercubes, one obtains\cite{smit82}
\begin{eqnarray}
\Gamma_{\mu\nu\alpha}(p,q)&=&\sum^{16}_{i=1}\Gamma^i_{\mu\nu\alpha}(p,q)
\nonumber\\
\Gamma^i_{\mu\nu\alpha}(p,q)&=&\Gamma^{i(\circ)}_{\mu\nu\alpha}(0)+
p_\rho\Gamma^{i(1)}_{\mu\nu\alpha ,\rho}(0)
+q_\rho\Gamma^{i(1)}_{\mu\nu\alpha ,\rho}(0)\nonumber\\
&+&\left(1-|_\circ - p_\rho|_\circ\partial_\rho - q_\rho|_\circ\partial_\rho
\right)
\Gamma^{con}_{\mu\nu\alpha}(p,q,m_i),
\label{smit1}
\end{eqnarray}
where $\Gamma^{con}_{\mu\nu\alpha}(p,q,m_i)$ can be evaluated in some continuum
regularizations. As for the 15 contributions of Dirac doublers
($i=2\cdot\cdot\cdot 15$), the first three terms in eq.(\ref{smit1}) 
are zero for the vector-like
Ward identity (\ref{gward}). The non-vanishing contributions are the same as the
15 copies of the $SU(2)$ vector-like gauge theory of massive Dirac fermions.
These contributions are gauge-invariant and finite (as $m_i\sim O(a^{-1})$),
thus, disassociate from the gauge anomaly. 

The non-trivial contribution from the
chiral mode in the hypercube $\Omega =[-\epsilon,\epsilon]^4$ is given by
\begin{eqnarray}
\Gamma^{i=1}_{\mu\nu\alpha}(p,q)\!&\!=\!&\!\int_\Omega\!
{d^4k\over(2\pi)^4}\tr\!\left[S(k\!+\!{p\over2})\Gamma_\mu(k)S(k\!-\!{p\over2})
\Gamma_\nu(k\!-\!{p\!+\!q\over2})S(k\!-\!{p\over2}\!-\!q)\Gamma_\alpha(k\!-\!
{q\over2})\right]\nonumber\\
&&+(\nu\leftrightarrow\alpha),
\label{tri}
\end{eqnarray}
where the propagator $S(p)$ and vertex $\Gamma_\mu(k)$ are given by
eqs.(\ref{sf},\ref{normal}). Other contributions containing anomalous vertices
$(\psi\bar\psi AA, \psi\bar\psi AAA)$ vanish within 
the hypercube $\Omega =[-\epsilon,\epsilon]^4$. We evaluate eq.(\ref{tri})
by the Pauli Villars regularization\footnote{The number of Pauli-Villars massive fermionic 
regulators is not infinite, which is different from ref.\cite{slanov}.} in the continuum, which certainly violates chiral gauge 
symmetry and is linearly divergent at the scale of $O(\epsilon)$.  As a result, modulo 
possible finite local counterterms, we obtain the consistent gauge anomaly 
for the non-abelian chiral gauge theories as the continuum one: 
\begin{equation}
\delta_g\Gamma(A')=-{ig^2\over24\pi^2}\int d^4x
\epsilon^{\alpha\beta\mu\nu}\tr\theta_a(x)\tau_a\partial_\nu 
\left[A_\alpha(x)\left(\partial_\beta A_\mu+{ig\over2}A_\beta (x)A_\mu(x)
\right)\right],
\label{anomaly}
\end{equation}
where the gauge field $A_\mu={\tau^a\over2} A^a_\mu$. 
The $SU_L(2)$ chiral gauge theory is anomaly-free for
$\tr(\tau^a,\{\tau^b,\tau^c\})=0$, as if an appropriate anomaly-free fermion content in
the group space. The gauge current,
\begin{equation}
J^a_\mu=i\bar\psi_L\gamma^\mu{\tau^a\over2}\psi_L={\delta\Gamma(A)\over \delta
A_\mu^a(x)}\hskip0.5cm \partial^\mu J^a_\mu=0
\label{conser}
\end{equation}
is covariantly conserved and gauge invariant. It must be pointed out and emphasized
that  in the hypercube $\Omega =[-\epsilon,\epsilon]^4$ we actually adopt a continuum (Pauli-Villars) regularization, which 
explicitly violates chiral gauge symmetries at the scale of $O(\epsilon)$, which is much smaller than the lattice scale $O(\pi/ a)$, for evaluating these 
anomalous terms like (\ref{anomaly}). The non-renormalization theorem of gauge anomaly 
guarantees that the resulted gauge anomaly is independent of any explicit breaking of chiral gauge symmetries at the scale of $O(\epsilon)$ and free from hight-order corrections. 

It seems surprising and impossible that we start from
a gauge symmetric action (\ref{action}) and fermion-field measure at the lattice scale, we end up with 
the correct form of the gauge anomaly (\ref{anomaly}).
Because, one normally claim that the anomaly has to come from the explicit breaking
of the chiral gauge symmetry in a regularized action (e.g., a Wilson term) at the lattice scale. 
This statement is indeed correct if regularized actions are
exactly local and bilinear in fermion fields, since this is nothing but what the ``no-go" theorem asserts. However, this is not correct in general. The general statement should be that the most essential and intrinsic {\it raison d'\^etre} of producing the
correct gauge anomaly in the lattice regularization is ``decoupling doublers'' rather than 
``explicitly breaking of chiral gauge symmetries at the lattice scale''.

In order to clarify and understant this general statement, let us first briefly review the relationship between doublers and chiral gauge anomalies in the lattice regularization. A most subtle property of the naive lattice regularization  of chiral gauge theories is the
appearance of 16 fermion zero modes. The gauge anomalies produced by these 16 fermion zero modes cancel each other. As a result, chiral gauge symmetries are exactly preserved at the lattice 
scale not only in the
naive lattice action but also lattice fermion-field  measure. This lattice fermion-field measure relates to the {\it finite} number of fermion-states (up to the lattice scale) of the vacuum of chiral gauge theories regularized by the lattice regularization for the finite lattice spacing $a$. While the anomalous currents of massless chiral 
fermions, that carry the fermion-states with definite chiral charges, flow into or out from the lattice regularized vacuum.
Since the number
of fermion-states of the lattice regularized vacuum is finite and these fermion-states are fully occupied, the total net  chiral charges carried by anomalous currents of massless chiral fermions must zero in the lattice regularized vacuum. Otherwise 
the total number of fermion-states of the lattice regularized vacuum would not be finite. 
This is the reason for the occurrence of 16-modes in 
4-dimension, the cancelation of chiral gauge anomalies produced by each chiral mode with
definite axial charge $Q_5$\cite{smit82}, and the chiral gauge symmetry is perfectly preserved. 
This subtle property is in fact an important merit of the lattice regularization. Contrastively, the number of fermion-states of the vacuum in other continuum 
regularization schemes is not exactly finite. The infinite number of hight-energy fermion-states of the vacuum in the continuum is only exponentially suppressed by explicit chiral-symmetry-violating terms in those continuum 
regularization schemes. In order to maintain this merit of the lattice regularization, doublers should be decoupled in a gauge invariant way rather than an explicit gauge variant way. 

In order to obtain the correct anomaly (\ref{anomaly}) in the lattice regularization, we obviously need to decouple extra doublers. If we adopt a local bilinear action to decouple
doublers, we must either explicitly break chiral gauge symmetries at the lattice scale $O(1/a)$ or 
give up the exactly locality as required by the ``no-go" theorem.  However, on the other hand, we run
into the dilemma that the gauge anomaly (\ref{anomaly}), obtained from an explicity breaking, is independent of any explicitly
breaking parameters(scale) (e.g., the Wilson parameter $r$($r/a$)), which is consistent with the non-renormalization theorem of gauge anomalies.  
In this sense, the explicity breaking at the lattice scale leading to the gauge anomaly is just a
superficial artifact in bilinear fermion actions. 

If we give up the bilinearity of regularized
actions in fermion fields and turn to our model and scenario with the exact chiral-gauge symmetry, the 15 doublers are decoupled as massive Dirac fermions that are vectorlike-gauge symmetric (\ref{gward}). Thus, they decouple from the gauge anomaly. Only the gauge anomaly associated with the normal (chiral) mode of the $\psi^i_L$ is
left and is the same as the continuum one, provided the right-handed three-fermion state $\Psi^i_R$ dissolves to the three-fermion-cut 
${\cal C}[\Psi^i_R]$ in the low-energy scale $\epsilon$ (\ref{threshold}). It would be otherwise that the massless right-handed three-fermion state $\Psi^i_R$ gives rise to a gauge anomaly exactly eliminating the gauge anomaly (\ref{anomaly}) associated to the massless left-handed fermion $\psi_L^i(x)$.
We still need to understand how the continuous states (virtual states) of the three-fermion cutes at the scale $\epsilon$ fill up the lattice regularized vacuum.

However, to be consistent with the manifest chiral gauge symmetry of the regularized theory
(action (\ref{action})) and fermionic measure,
the gauge anomaly (\ref{anomaly}) must be canceled within the fermion content of the theory. Otherwise, the vectorlike spectrum of fermion zero modes must appear, either doublers do not decouple or the right-handed 
three-fermion-state $\Psi^i_R$ does not dissolve into its cut and becomes a massless right-handed particle in the low-energy.
From this point of view, we see the anomaly-cancelation by the fermion content is 
a necessary condition\footnote{It is obvious not a sufficient condition.} for this 
scenario to work, in particular, for the Ward identities (\ref{gward}) and (\ref{glward}) 
to be valid. 

Before ending this section, we would like in particularly to discuss the residual breakings $R(a)$ of chiral gauge symmetries \cite{gs99,testa99} after the gauge anomaly (\ref{anomaly}) is canceled  $\delta_g\Gamma(A)=0$,
\begin{equation}
\delta\Gamma(A)=\delta_g\Gamma(A)+R(a).
\label{reb}
\end{equation}
If the gauge anomaly is induced by explicit breakings of the chiral gauge symmetry  at the lattice scale, there must be the residual breakings $R(a)$ of the gauge symmetry at the lattice scale. Normally, given an explicity breaking of the chiral gauge symmetry at the lattice scale, we can perturbatively compute the gauge anomaly (\ref{anomaly}) and residual breakings of gauge symmetry by a small and smooth background of the external classical gauge field. Here by a smooth and small background, we indicate that the correlations of the gauge field  are much larger than the lattice spacing $O(a)$ and the fluctuations of the gauge field  are much smaller than the lattice scale $O(1/a)$.  In such a background of the external gauge field, except the non-local gauge anomaly that is eliminated, the residual breakings of gauge symmetry are local and high dimension irrelevant operators at the lattice scale. However, these residual breakings of gauge symmetry could turn out to be relevant for the large and no smooth fluctuations of the longitudinal gauge field at the order of the lattice scale, and attempted lattice chiral gauge theories are jeopardized by breaking chiral gauge symmetries. 

While, in our scenario, no smooth  ($O(a)$) and the non-perturbatively large fluctuations ($O(1/a)$) of  the longitudinal gauge field at the order of the lattice scale are fully under controlled by the chiral gauge symmetric and vectorlike spectra of 15 non-degenerated massive Dirac fermions. In the background of the external gauge field (the transverse and longitudinal components) with small fluctuations ($O(\epsilon)$) and smooth correlations ($O(1/\epsilon)$) at the order of the scale $\epsilon$ (\ref{threshold}), we compute the vacuum functional and obtain the gauge anomaly. The residual breakings $R(\epsilon)$ of gauge symmetry, which comes out as the companions of the genuine gauge anomaly, are due to the explicity chiral symmetry breaking introduced by a continuum regularization scheme at the scale $\epsilon$ (\ref{threshold}), rather than at the lattice scale. Since the non-local gauge anomaly is canceled, chiral gauge symmetry is exact in the scaling region 
${\cal A}$ (\ref{segment}) and the vacuum functional is determined up to local counterterms, we can subtract these residual breakings $R(\epsilon)$ of gauge symmetry away by adding appropriate local counterterms, in order to achieve an asymptotically chiral gauge field theory in the continuum limit.  This is the same as the procedures in the normal renormalization prescription of quantum field theories in the continuum.

\section{The anomalous fermion-flavour singlet current}

The non-conservation of fermion numbers is an important feature of the Standard
Model. A successful regularization of chiral gauge theories should give this
feature in the continuum limit. In the Eichten-Preskills model of 
multifermion couplings for the $SU(5)$ and $SO(10)$ theories, inspired by the 
origination of the axial anomaly in the
lattice QCD, it was suggested that the anomalous global current should be
originated from the explicit breaking of the global symmetry at the lattice scale. In the context of the standard model, an elegant four-fermion interacting vertex 
of explicitly violating fermion number (B-L) was introduced\cite{mc}. It is expected that the correct 
anomalous fermion-flavour singlet currents and violating fermion numbers should consistently 
be obtained. Nevertheless, we need to do explicit calculations to obtain the anomalous 
fermion-flavour singlet current in the standard model and $SU(5),SO(10)$ unification theories.

In this section, within the scenario presented in previous sections, we show that this fermion-flavour singlet anomaly
can be consistently obtained from the explicit chiral gauge and fermion-flavour symmetric action (\ref{action}) and fermion-field measure at the lattice scale. We first present the explicit calculations to obtain the correct fermion-flavour singlet  anomaly and then discuss the consistency and reason for achieving this anomaly in our scenario.

Our action (\ref{action}) processes the $U_L(1)$ and $U_R(1)$ global chiral 
symmetries. At the lattice scale, the action is invariant under the following 
global transformations:  
\begin{equation}
\psi_L^i\rightarrow e^{i\theta_L} \psi_L^i\hskip1cm
\chi_R\rightarrow e^{i\theta_R}\chi_R,
\label{dd}
\end{equation}
where $\theta_{L,R}$ are the $U_{L,R}(1)$ chiral phases.
These global symmetries lead to the conservation
of the singlet chiral fermion currents,
\begin{eqnarray}
\partial_\mu j^\mu_L(x)&=&0,\hskip0.5cm
j^\mu_L=i\bar\psi_L^i\gamma^\mu\psi^i_L\label{lcu}\\
\partial_\mu j^\mu_R(x)&=&0,\hskip0.5cm
j^\mu_R=i\bar\chi_R\gamma^\mu\chi_R +O(a^2),
\label{rcu}
\end{eqnarray}
where $j^\mu_{L,R}$  are Noether currents and $\partial_\mu$ is the derivative on the lattice. Eqs.(\ref{lcu},\ref{rcu}) correspond to the conservation
of the fermion numbers of $\psi_L^i$ and $\chi_R$. However, as we know, eq.(\ref{lcu}) should be anomalous. 

In order to see whether the conservations of the currents are violated when the external
chiral gauge field is coupled to chiral fermions, we consider the source
currents $\langle j^\mu_L(x)\rangle$ and $\langle j^\mu_R(x)\rangle$ defined as
\begin{eqnarray}
\langle j^\mu_{L}(x)\rangle &=&{\delta_L \Gamma(A')\over\delta 
V^{L}_\mu(x)};
\hskip1cm \delta V^{L}_\mu(x)= -\partial_\mu\theta_{L}(x);\label{sourcel}\\
\langle j^\mu_{R}(x)\rangle &=&{\delta_R \Gamma(A')\over\delta 
V^{R}_\mu(x)};
\hskip1cm \delta V^{R}_\mu(x)= -\partial_\mu\theta_{R}(x),
\label{sourcer}
\end{eqnarray}
where $\Gamma(A')$ is the vacuum functional. Under the variations $\delta_L$ and $\delta_R$ of the $U_{L,R}(1)$-phases
$\theta_{L}(x)$ and $\theta_{R}(x)$,
the vacuum functional $\Gamma$ is transformed up to $O(\theta_{L})$ and
$O(\theta_{R})$,
\begin{eqnarray}
\delta_{L}\Gamma &=&\int d^4x\delta V^{L}_\mu(x)\langle j^\mu_{L}(x)
\rangle=\int d^4x \theta_{L}(x)\partial_\mu
\langle j^\mu_{L}(x)\rangle,\label{deltall}\\
\delta_{R}\Gamma &=&\int d^4x\delta V^{R}_\mu(x)\langle j^\mu_{R}(x)
\rangle=\int d^4x \theta_{R}(x)\partial_\mu
\langle j^\mu_{R}(x)\rangle,
\label{deltarr}
\end{eqnarray}
where
\begin{eqnarray}
\delta_{R}\Gamma &=& \Gamma(A'_\mu+\delta V^{R}_\mu(x))-\Gamma(A'_\mu),
\label{variationr}\\
\delta_{L}\Gamma &=& \Gamma(A'_\mu+\delta V^{L}_\mu(x))-\Gamma(A'_\mu).
\label{variationl}
\end{eqnarray}
In our action (\ref{action}), the $\chi_R$ does not directly couple to the external
gauge field and this decoupling strictly holds due to the Ward identity
(\ref{wa}). Thus, the fact that the $\langle j^\mu_R(x)\rangle$ defined formally in
eq.(\ref{sourcer}) does not couple to the external gauge field $A_\mu$, i.e., in eq.(\ref{variationr})
\begin{equation}
\delta_R\Gamma(A')=0,
\label{gs}
\end{equation}
leads to two direct consequences. One is the exact conservation of the source current $\langle j^\mu_R(x) \rangle$
\begin{equation}
\partial_\mu\langle j^\mu_R(x) \rangle=0,
\label{2}
\end{equation}
from (\ref{deltarr}). Another the exact gauge invariant source current $\langle j^\mu_R(x)\rangle$,
\begin{equation}
\delta_g\langle j^\mu_R(x)\rangle=\delta_g{\delta_R \Gamma(A')\over\delta 
V^{R}_\mu(x)}=0,
\label{1}
\end{equation}
where $\delta_g$ is a gauge variation.
The facts that the Ward identity of the $\chi_R$-shift-symmetry is not violated by the gauge interaction in the action (\ref{action}) and the gauge field is completely decoupled from 
the neutral sector (\ref{wa}), are extremely crucial to the conservation
of the right-handed fermion numbers (\ref{2}). The same reason leads to the exact conservation 
of the neutral and composite left-handed current $J^{\mu,n}_L=i\bar\Psi^n_L\gamma_\mu
\Psi^n_L$, i.e.,  $\partial_\mu J^{\mu,n}_L=0$.

As discussed in section 5, under a gauge variation $\delta_g$ , in
general we have, 
\begin{equation}
\delta_g\Gamma_{\mu\nu\alpha}(p,q)\not=0,\hskip1cm
\delta_g\Gamma^{(n)}_{\{\mu\}}=0\hskip0.5cm (n\not=3),
\label{nm}
\end{equation}
where the vacuum functional $\Gamma(A')$ is just the same as the continuum counterpart 
up to some gauge-invariant finite terms. In the anomaly-free $SU_L(2)$ case, i.e.,
$\delta_g\Gamma(A)=0$, one may conclude that the source current $\langle j^\mu_L(x)
\rangle$ defined in eq.(\ref{sourcel}) is gauge invariant, 
\begin{equation}
\delta_g\langle j^\mu_L(x)
\rangle=\delta_g{\delta_L \Gamma(A)\over\delta 
V^L_\mu(x)}={\delta_L\delta_g\Gamma(A) \over\delta V^L_\mu(x)}=0,
\label{gauin}
\end{equation}
and the variation (\ref{variationl}) vanishes, i.e., $\delta_L \Gamma=0$, 
leading to  $\partial_\mu\langle j^\mu_L(x)
\rangle=0$ by eq.(\ref{deltall}) . However, these are not true. The order of the differentiations $\delta_g$ and 
$\delta_L$ can not be exchanged and $\delta_L \Gamma\not=0$ because of the mixing anomaly. 

We know that in our action (\ref{action}), the
left-handed variation
\begin{equation}
\delta V^L_\mu(x)= - \partial_\mu\theta_L(x),
\label{vl}
\end{equation}
can be considered as a commuting $U_L(1)$ factor in the $SU_L(2)$ chiral 
gauge group, i.e.,
\begin{equation}
\tilde A_\mu=A_\mu+V_\mu^L,\hskip1cm (A_\mu={\tau^a\over2} A^a_\mu).
\label{va}
\end{equation}
Actually,
this is an $SU_L(2)\otimes U_L(1)$ chiral gauge group and there is a mixing
anomaly\cite{preskill91}, 
\begin{eqnarray}
\delta_L\Gamma &=& C_1{ig^2\over32\pi^2}\int d^4x
\theta_L\tr\left(F_{\mu\nu}\tilde F^{\mu\nu}\right),\label{arbi1}\\
\delta_g\Gamma&=& C_2{ig\over16\pi^2}\int d^4x
\tilde F^{\mu\nu}_1\tr\left(\theta_g\partial_\mu A_\nu\right),
\label{arbi}
\end{eqnarray}
where $\theta_g=\theta^a_g\tau_a$ is the $SU_L(2)$-transformation parameter, $C_1,C_2$ are
arbitrary constants with $(C_1+C_2=1)$, and 
\begin{eqnarray}
F^{\mu\nu}&=&\partial^\mu A^\nu-\partial^\nu A^\mu,\nonumber\\
F^{\mu\nu}_1&=&\partial^\mu V^{L\nu}-\partial^\nu V^{L\mu} .
\label{f1}
\end{eqnarray}
The reason is that one of the Pauli
matrices $\tau^a / 2$ in the triangle graph is replaced by the generator
(identity) of the $U_L(1)$, i.e., the $U_L(1)$ global current, therefore 
the vanishing of the $SU_L(2)$ anomaly for
$\tr(\tau^a,\{\tau^b,\tau^c\})=0$ is no longer true. Note that in 
eqs.(\ref{arbi1},\ref{arbi}), we only consider
the triangle diagram $(n=3)$, since 
\begin{equation}
\delta_L\Gamma^{(n)}_{\{\mu\}}=0,\hskip0.3cm\delta_g\Gamma^{(n)}_{\{\mu\}}=0\hskip0.3cm 
(n\not=3)
\label{3}
\end{equation}
for $\Gamma^{(n)}_{\{\mu\}} (n\not=3)$ being gauge-invariant, as we discussed
in section 5. 

The mixing anomaly (\ref{arbi}) has arbitrariness $C_1$ and $C_2$,
which arise because the triangle graphs with one insertion of the $U_L(1)$
global current determine the $\Gamma_{\mu\nu\alpha}(A')$ up to local
counterterms. As the Feynman diagrams determine the vacuum functional $\Gamma
(A')$ only up to an arbitrary choice of local counterterms, we are allowed to
add local counterterms into the vacuum functional 
\begin{equation}
\Gamma'(A')=\Gamma(A')+\Gamma_{c.t.}(A'),
\label{reg}
\end{equation}
which is equivalent to the
re-definition of the chiral fermion current defined by eq.(\ref{sourcel}), 
\begin{equation}
\langle j^{'\mu}_L\rangle=\langle j^{'\mu}_L\rangle+\langle j^{'\mu}_{L,c.t.}\rangle.
\label{regj}
\end{equation}

Due to the fact that in our scenario the vacuum functional $\Gamma(A')$ we obtained for the 
$SU_L(2)$ case is free from the non-local gauge anomaly and local
gauge-symmetry-breaking terms, 
$\delta_g\Gamma$ (\ref{arbi}) must vanish and the
arbitrariness in eq.(\ref{arbi1},\ref{arbi}) can be fixed, 
\begin{equation}
C_1=1,\hskip0.5cm C_2=0
\label{c1c2}
\end{equation}
by choosing an
adequate local counterterm. As a result, the vacuum functional and the
left-handed current are gauge invariant,
\begin{equation}
\delta_g\Gamma'(A')=0,\hskip1cm \delta_g
\langle j^{'\mu}_L\rangle=0.
\label{pp}
\end{equation}
From eqs.(\ref{deltall},\ref{arbi1}), we obtain
\begin{equation}
\delta_L\Gamma'={ig^2\over32\pi^2}\int d^4x\theta_L\tr\left(F_{\mu\nu}\tilde 
F^{\mu\nu}\right);\hskip0.5cm\partial_\mu \langle j^{'\mu}_L\rangle =
{ig^2\over32\pi^2}\tr\left(F_{\mu\nu}\tilde F^{\mu\nu}\right).
\label{fer}
\end{equation}
This is just the desired result, showing the left-handed fermion number is
violated by the $SU(2)$ instanton effect, which attributes to the topological configuration of the gauge field carrying fermion numbers. By the definition  $ \langle j^{\mu}_5\rangle = \langle j^{'\mu}_L\rangle- \langle j^{\mu}_R\rangle$
and $\partial_\mu \langle j^{\mu}_R\rangle =0$ (\ref{2}), we obtain,
\begin{equation}
\partial_\mu \langle j^{\mu}_5\rangle =
{ig^2\over32\pi^2}\tr\left(F_{\mu\nu}\tilde F^{\mu\nu}\right).
\label{fer5}
\end{equation}
  
We emphasize crucial points for achieving the correct form of fermion-flavour singlet anomaly (\ref{fer})
up to gauge invariant local counterterms in this scenario. The first is the right-handed
composite three-fermion-state $\Psi^i_R(x)$ dissolving into its three-fermion-cut $ {\cal C }\Psi^i_R(x)$, in another word, no massless right-handed
composite three-fermion-state $\Psi^i_R(x)$ exists in the low energy so that the correct from of the gauge anomaly is obtained (\ref{anomaly}). The second is the decoupling of the right-handed massless fermion $\chi_R$ from the gauge field,  which leads to $\partial_\mu \langle j^{\mu}_R\rangle=0$. The third is exact chiral gauge symmetry, i.e., no chiral gauge-symmetry-breakings, and the vanishing of the gauge anomaly so as to have the choice (\ref{c1c2}) up to local gauge invariant counterterms.  The fourth is the correct gauge anomaly obtained in the continuum regularization scheme which explicitly breaks the chiral symmetry at the scale $O(\epsilon)$ being much smaller than the lattice scale. In fact, instead of using the formula of the mixing anomaly (\ref{arbi1}, \ref{arbi}), we can directly compute the triangle diagrams (\ref{tri}) with the left-handed fermion-flavour singlet current insertions in the continuum Pauli-Villars regularization scheme for the hypercube $\Omega =[-\epsilon,\epsilon]^4$, to achieve the correct fermion-flavour singlet anomaly (\ref{fer}). However, the mixing anomaly let us have a clear connection between the gauge anomaly-free, chiral gauge symmetry and the fermion-flavour singlet anomaly.

We turn to the discussions  why we obtain the correct fermion-flavour 
$U_L(1)$-anomaly from the $U_L(1)$ symmetric action (\ref{action}) at the lattice scale.
This question arises  because of our
knowledge of the lattice QCD where the axial current anomaly, 
\begin{equation}
\partial_\mu j^\mu_5=
{ig^2\over16\pi^2}\tr\left(F_{\mu\nu}\tilde F^{\mu\nu}\right)
\label{aa}
\end{equation}
is due to the flavour $SU_L(3)\otimes SU_R(3)$ asymmetric Wilson term at the lattice scale.
Certainly, since we start from a global symmetric action and no spontaneous symmetry breaking occurs, the relevant operators in the scaling region must have the same symmetries as they are at the lattice scale. No way to generate 
chirally asymmetric relevant operators by its own dynamics to produce the fermion-flavour singlet anomaly. All what we have done for producing the fermion-flavour singlet anomaly is introducing an explicit and soft chiral symmetry breaking at the scale $O(\epsilon)$ consistently with the dissolving scale of the right-handed three-fermion state $\Psi^i_R$.
The reason that we obtain the correct fermion-flavour singlet anomaly is very analogous to that we obtain the correct gauge anomaly discussed in the end of section 5. 

{\it A priori}, we have no any dynamical reasons to expect that the non-conservation
of fermion number must be due to the explicit chiral symmetry breaking of the global $U_L(1)$ symmetry {\it at
the lattice scale}. Further, the fact that the resulted fermion-flavour singlet current anomaly is independent of the 
explicit-symmetry breaking parameters implies that 
the explicit-symmetry-breaking is not necessary at the lattice scale, and can be at other scales, for instance $\epsilon$ (\ref{threshold}), much smaller than the lattice scale. This is also consistent with the non-renormalization theorem of anomalies. However, in order to achieve the 
fermion-flavour singlet anomaly by the explicit chiral symmetry breaking at a much softer scale than the lattice scale, we have to pay price as required by the ``no-go" theorem of Nielsen-Ninomiya, either relaxing the exact locality of the theory, which is fundamental property of quantum field theories, or giving up the bilinear construction of fermionic action and running to multifermion interactions at the lattice scale. The appearance of two scales in our scenario, the lattice scale and the dissolving scale $\epsilon$ (\ref{threshold}), is reminiscence of the two cutoff approach to chiral gauge theories on the lattice\cite{thooft}.

In fact, the fermion-flavour singlet
anomaly (\ref{fer}) must disappear as the external gauge field is turned off. The conservations
of the fermion-flavour singlet currents (\ref{2},\ref{fer}) must be related to the conservations of Noether currents (\ref{lcu},\ref{rcu}) coming from the explicit global $U_L(1)$ and $U_R(1)$
symmetries of the action (\ref{action}) in the absence of gauge field at the lattice scale(tree-level).
Otherwise, we would run to the dilemma that when the external gauge field (or the topological configuration of the gauge field) is turned off and the fermion-flavour singlet current (\ref{fer}) becomes conserved, however, on the other hand, the corresponding Noether current (\ref{lcu}) is not conserved due to the explicit $U_L(1)$ chiral-symmetry-breaking introduced  at the lattice scale (tree level).   

In our scenario, we still need to have more intuitive understanding of how this anomalous fermion number, contributed by the presence of the topological configuration of the gauge field, flows into and emerges
out from the regularized vacuum and how the continuous states (virtual states)  of the three-fermion cutes at the scale $\epsilon$ relate to the anomalous fermion number flows.

The multifermion couplings we considered in this paper is particular one in the context of the standard model. Taking the doublet $\psi_L^i$
and spectator $\chi_R$ as the $SU_L(2)$ doublet and  right-handed neutrino in the leptonic sector of the standard model, we have the gauge symmetric multifermion couplings in the action (\ref{action}). However, in the fermion content of the standard model, we certainly have the possibilities of fermion-number violation, but chiral gauge symmetric multifermion couplings at the lattice scale. The nice examples are given in refs.\cite{ep,mc}. If a scaling region with the desired gauge symmetric spectrum in the low-energy can be achieved, these fermion-number violating multifermion couplings would turn out to be relevant operators in such a scaling region to give the $B-L$ fermion number violation. It could be  the Nature choice.

\section{Conclusions and discussions}

In this paper, we present the analysis and scenario of the chiral fermion spectrum and 1PI interacting vertices
between fermions and gauge field in the low-energy scaling region (\ref{segment}) of the model (\ref{action}) proposed in ref.\cite{xue97,xue99} for chiral gauged fermions on the lattice. In addition, we show how the gauge anomaly and fermion-flavour singlet anomaly are correctly produced in such a scenario. This shows that an asymptotic chiral gauge theory in the continuum limit can be realized in the low-energy scaling region (\ref{segment}) of our model (\ref{action}). We conclude that our model and scenario provide a plausible solution to the long-standing problem of chiral gauged fermion on a lattice. It is very inviting that numerical computations and other techniques to verify our scenario, in particular, the phenomena of no hard spontaneous symmetry breaking at the lattice scale and the intermediate energy-threshold $\epsilon$ (\ref{threshold}) where the three-fermion-state $\Psi_R^i(x)$ dissolve into its corresponding cut ${\cal C} [\Psi_R^i(x)]$. 
It is also very necessary to analyze an analogous model in 1+1 dimensions\cite{xue00} so as to make our conclusion be more convincible. In addition, the consistency of our model in the scaling region regarding the $SU_L(2)$ global Witten anomaly\cite{witten} is still open question. These are subject to the future work. 

In this section, we wish to make a very general and brief discussion on any possible relationships between the multifermion coupling and bilinear fermion coupling approaches for anomaly-free chiral gauge theories on the lattice. 
In both bilinear fermion and multifermion coupling models, extra fermionic species must be decoupled and right-handed and left-handed fermionic species must couple in some ways to have anomalies. The couplings between right- and left-handed fermionic species appear either in the action or in the fermionic measure. 

The Wilson fermion is exact local (in the range of the lattice spacing) and  right- and left-handed fermionic species couple at the lattice scale. Doublers are very massive and decoupled. As required by the ``no-go" theorem, the residual breakings of the gauge symmetry are at the lattice scale. These residual breakings of the chiral gauge symmetry can be eliminated by adding and fine-tuning appropriate counterterms so as to enforce the Ward identities associated to exact chiral gauge symmetries at the continuum limit\cite{rome}. 

Contrasting with the Wilson fermion, the regularization of the fermion sector adopted by the ``overlap"\cite{nerev} and L\"uscher\cite{lurev} approaches use the Ginsparg-Wilson equation\cite{gw} that was obtained from the renormalization group equation. Owing to the Dirac operator satisfying the Ginsparg-Wilson equation, the residual breakings of the gauge symmetry are reduced\cite{gs99} and supposed to be eliminated by either average over gauge configurations or adding local counterterms in order to preserve exact chiral gauge symmetries. In the L\"uscher\cite{lurev} approach for the abelian gauge theory, all residual breakings of the gauge symmetry including the gauge anomaly can be rewritten as a total divergence for its topological nature, thus they are eliminated by redefining the gauge current\footnote{Private conmmunication with L\"uscher.} for the finite lattice spacing and without fine-tuning. In order to completely decouple extra fermion species, as required by the ``no-go" theorem, the approaches relax the exact locality to the locality whose range extends  to a few lattice spacings with an exponential tail. 

While in the models of multifermion couplings, the couplings of right-handed and left-handed fermions can be made exactly local and chiral gauge symmetric at the lattice scale. However, as discussed at the end of section 2, we have to find a peculiar multifermion coupling and a scaling region desired for the low energy. The hard spontaneous symmetry breaking is absolutely not tolerated so that residual breakings of gauge symmetries are not at the lattice scale. The strong coupling at the high energy is needed and three-fermion cut must be realized at the low-energy so as to decouple extra fermion species with ``wrong" chirality. 

Taking our action (\ref{action}) as an example, we can formally integrate away the spectator field $\chi_R$ and obtain the effective Dirac action  bilinearly in the fermion field $\psi_L^i$. Such an effective Dirac action is obviously not exactly local. It is worthwhile to examine whether such an effective Dirac operator could be the solution to the Ginsparg-Wilson equation in the sense of the renormalization group invariance. Most importantly, we need to show in the scaling region ${\cal A}$ (\ref{segment}) for the low-energy, whether the relevant spectra and operators induced from the multifermion couplings(high dimension operators) at the lattice scale are in the same universal class with the solutions to the Ginsparg-Wilson equation, in the view of the renormalization group invariance. 

In fact, the recent successful progress based on the Ginsparg-Wilson equation strongly implies the existence of the scaling region \ref{segment}) for exactly chiral-gauge symmetric theories in the low-energy, obtained from our model (\ref{action}). This has been generally believed to be impossible. 
The studies of appropriate multifermion couplings at the lattice scale and desired scaling region are highly deserved, since they could be Nature's choice for chiral gauge theories, e.g., the Standard Model, at the high-energy.

\section{Acknowledgements}

I thank to the organizers of the winter workshop at Benasaque Spain (January 2000), where the manuscript of paper was completed.

\end{document}